\documentclass[prb,superscriptaddress,twocolumn,showkeys]{revtex4}
\usepackage{graphicx}
\usepackage{hyperref}
\usepackage{amsmath}
\usepackage{amssymb}
\usepackage{bm}
\usepackage{units}
\usepackage{ulem}

\usepackage{mathtools}
\usepackage{mathdots}

\usepackage[dvipsnames]{xcolor}
\usepackage[utf8]{inputenc}  
\usepackage{fix-cm}  % this package allows large \fontsize
\usepackage{tikz}    % this is for graphics. e.g. rectangle on title page
\usepackage{array}
\usetikzlibrary{arrows}
\usetikzlibrary { decorations.pathmorphing, decorations.pathreplacing, decorations.shapes, }
\usepackage{tikz-cd} 
\usepackage{tkz-euclide}
\usepackage{circuitikz}
\usepgfmodule{shapes}
\usetikzlibrary{decorations.pathmorphing}
\usetikzlibrary{backgrounds, arrows,calc,shapes,decorations.pathreplacing, automata,positioning}
\usepackage{cancel} 
\usepackage{graphicx}
\usepackage{array}
\usetikzlibrary{decorations.markings}

\usetikzlibrary{decorations.pathreplacing}
\usetikzlibrary{shapes.misc}

\def\SQCD{SQCD$_3$}

\tikzset{cross/.style={cross out, draw=black, minimum size=5*(#1-\pgflinewidth), inner sep=0pt, outer sep=0pt},
cross/.default={2pt}}

\tikzset{
    arrowMe/.style={
        postaction=decorate,
        decoration={
            markings,
            mark=at position .5 with {\arrow[thick]{#1}}
        }
    }
}
\pgfarrowsdeclaredouble{<<s}{>>s}{stealth}{stealth}%   double stealth

\tikzset{snake it/.style={decorate, decoration=snake}}

\tikzset{mid arrow/.style={postaction={decorate,decoration={
        markings,
        mark = at position .55 with {\arrow[#1]{Straight Barb[width=5pt]}}
      }}}}
      
\tikzset{mid arrowsm/.style={postaction={decorate,decoration={
        markings,
        mark = at position .55 with {\arrow[#1]{Straight Barb[width=3pt]}}
      }}}}
      
\tikzset{middx arrowsm/.style={postaction={decorate,decoration={
        markings,
        mark = at position .7 with {\arrow[#1]{Straight Barb[width=3pt]}}
      }}}}
      
\tikzset{midsx arrowsm/.style={postaction={decorate,decoration={
        markings,
        mark = at position .4 with {\arrow[#1]{Straight Barb[width=3pt]}}
      }}}}

\usetikzlibrary{external}
\usepackage{amsmath} % Used by equations
\usepackage{amssymb}
\usepackage{graphicx}
\graphicspath{ {images/} }
\usepackage{caption}
\usepackage{subcaption}
\newcommand{\INFNT}{Istituto Nazionale di Fisica Nucleare, Sezione di Trieste, Via Valerio 2, I-34127 Trieste, Italy}
\newcommand{\INFNM}{Istituto Nazionale di Fisica Nucleare,  Sezione di Milano, Via Celoria 16, I-20133 Milano, Italy}
\newcommand{\MIB}{Dipartimento di Fisica, Universita di Milano-Bicocca, Piazza della Scienza 3, I-20126 Milano, Italy}
\newcommand{\SISSA}{SISSA, Via Bonomea 265, I-34136 Trieste, Italy}

\begin{document}

\title{Planar Abelian Mirror Duals of $\mathcal{N}=2$ SQCD$_3$}

\author{Sergio Benvenuti}
\email{benve79@gmail.com}
\affiliation{\INFNT}
\author{Riccardo Comi}
\email{r.comi2@campus.unimib.it}
\affiliation{\MIB}
\affiliation{\INFNM}
\author{Sara Pasquetti}
\email{sara.pasquetti@gmail.com}
\affiliation{\MIB}
\affiliation{\INFNM}
\author{Gabriel Pedde Ungureanu}
\email{gpeddeun@sissa.it}
\affiliation{\INFNT}
\affiliation{\SISSA}
%\affiliation{\INFNT}
\author{Simone Rota}
\email{srota@sissa.it}
\affiliation{\INFNT}
\affiliation{\SISSA}
%\affiliation{\INFNT}
\author{Anant Shri}
\email{ashri@sissa.it}
\affiliation{\INFNT}
\affiliation{\SISSA}

\begin{abstract}
We propose an Abelian mirror dual for the $\mathcal{N}=2$ SQCD$_3$ that we obtain as real mass deformation of known $\mathcal{N}=4$ mirror pairs. 
We match the superconformal index and the $\mathbf{S}^3_b$ partition function, discuss the agreement of the moduli spaces, and provide a map of the gauge invariant operators and the global symmetries as evidence of this duality.
\end{abstract}

\maketitle

\section{Introduction}

We propose an \textit{Abelian} dual description for a family of $\mathcal{N}=2$  $SU(N)$ Chern-Simons (CS) \SQCD\cite{Boer_1997,Aharony_1997,Intriligator_2014}. The dual theory is \textit{planar}, in the sense that it is a quiver drawn on a plane, with a cubic superpotential term for each closed oriented loop, as in the $4d$ periodic quivers associated to the dimers of \cite{Hanany:2005ve, Franco:2005rj}
and, in addition, contains linear monopole superpotentials. Our dualities are obtained by 
real mass deformations of 3d $\mathcal{N}=4$ mirror pairs of theories\cite{Intriligator_1996,Hanany_1997} and display the exchange of topological and flavor symmetries characteristic of $3d$ mirror symmetry.

In this letter, we focus on the $SU(N)_k$ \SQCD\ with $F\geq2N$ fundamentals and CS level $k=\frac{F}{2} - N$ and discuss extensions of the analysis to larger $k$ via further real mass deformations.
We can extend our results to theories with more general flavor content, CS levels, and quivers with unitary gauge groups as discussed in forthcoming papers \cite{Benvenuti_2024, Benvenuti_2024a}. 
As an example, we discuss a duality between chiral and planar quivers obtained from the self-mirror $T[SU(N)]$ theory\cite{Gaiotto_2008}.

Our proposal is guided by the behavior under real mass deformations of the $\mathbf{S}^3_b$ partition function\cite{Kapustin:2009kz, Hama_2011}, which matches across our proposed duality, providing a non-trivial check of our claims. Details of this analysis will be given in  \cite{Benvenuti_2024, Benvenuti_2024a}.
In this letter, we support our proposals by matching the Superconformal Index\cite{Imamura_2011, Kapustin_2011} (SCI), the global symmetries, and the chiral rings of the dual theories.

%%%%%%%%%%%%%%%%%%%%%%%%%%%%%%%%%%%%%%%%%%%%%%%
\section{A Planar Abelian Dual for $SU(N)_{\frac{F}{2}-N}$ with $F$ fundamentals}\label{sec:1ststep}
\begin{figure}
\includegraphics[]{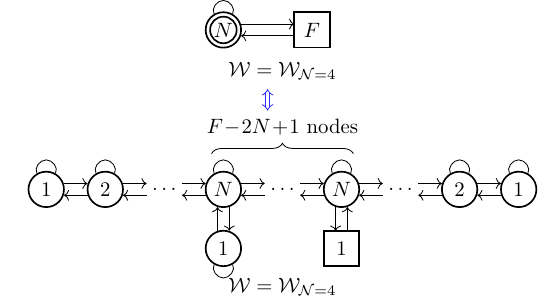}
\caption{$\mathcal{N}=4$ mirror duality for SQCD$_3$. Single/double circles correspond to $U$/$SU$ symmetry groups.}
\label{fig:N4}
\end{figure}

We start from the $\mathcal{N}=4$ mirror duality relating $SU(N)$ with $F \geq 2N$ flavors to a quiver with $F$ gauge nodes %\cite{Intriligator_1996, Hanany_1997} 
(Figure \ref{fig:N4}).
We deform the $\mathcal{N}=4$ electric theory,
breaking the $SU(2)\times SU(2)$ R-symmetry to $U(1)_R$\cite{Tong:2000ky} by turning on a real mass $m$ for a combination of the commutant of $U(1)_R$ and the baryonic symmetry. 
Under this deformation, in the vacuum where the real scalar has no VEV, only the $F$ fundamental fields remain massless while the adjoint and the $F$ antifundamental fields are massive and are integrated out generating a non-zero CS level and we obtain the $\mathcal{N}=2$ $SU(N)_{\frac{F}{2}-N}$ \SQCD\, with $F$ fundamental fields \footnote{The contribution to the shift in CS level is $+\frac{F}{2}$ for fermions in the anti-fundamental representation, and $-2 \times \frac{N}{2}$ for the fermions in the $\mathcal{N}=2$ adjoint chiral.}.

\begin{figure}[t!]
\includegraphics[]{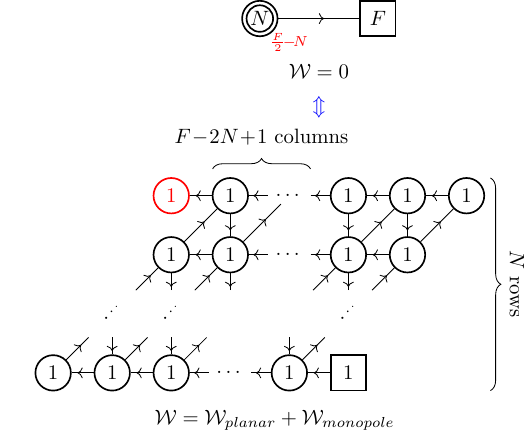}
\caption{Mirror-like duality for $\mathcal{N}=2$ \SQCD. }
\label{fig:N2}
\end{figure}

In the magnetic theory, under the same deformation,  something interesting happens - 
we propose that the magnetic vacuum corresponding to the electric theory is such that each gauge node $U(k)$ is Higgsed to its maximal torus $U(1)^k$ producing a column of $k$ black $U(1)$ gauge nodes as in Figure \ref{fig:N2}.
For a given node, the Higgs mechanism is triggered by
VEVs for the scalar $\sigma$ in the corresponding $\mathcal{N}=2$ vector multiplet. % trigger the Higgs mechanism. 
To reach this vacuum we move along the Coulomb Branch (CB), where $\langle\sigma\rangle=\text{diag} (\sigma_1,\sigma_2,\dots)$ with:
\begin{equation}\label{eq: VEV}
\sigma_i - \sigma_{i+1} = m
\end{equation}
and the chiral fields depicted in Figure \ref{fig:N2} remain massless.
We checked that this vacuum satisfies the F-term and D-term equations \cite{Intriligator_2014}. 

This claim is supported by  the careful analysis of the effect of the real mass deformation on the $\mathbf{S}_b^3$
partition function as in \cite{Fokko_2007,Benini_2011}.
The large mass limit produces a highly oscillating phase corresponding to contributions from massive fields and is sensitive to the Higgsing\cite{Aharony_2013}.
Assuming the Higgsing pattern described above, we checked that the  highly oscillating phases
%associated with the massive fields 
cancel between the electric and magnetic sides. This procedure implies the equality of the partition functions of the two theories in Figure \ref{fig:N2} and is a very non-trivial check of the duality. Details on these computations will appear in a forthcoming paper \cite{Benvenuti_2024}.

The mirror dual theory admits a Lagrangian description, all the gauge groups are Abelian and each arrow denotes a chiral field with charge $(+1,-1)$ under the two nodes it connects. Importantly, the theory also contains (mixed) CS interactions and a non-zero superpotential:
\begin{itemize}
    \item each black (red) gauge node carries a $-1$ ($-\frac{1}{2}$) CS coupling. The CS interactions are a consequence of  integrating out fermionic fields present in the $\mathcal{N}=4$ theory.
    \item For each vertical (non-vertical) arrow there is a $-1$ ($+1$) mixed CS coupling involving the two nodes connected by the arrow.
    \item $\mathcal{W}_{planar}$ are cubic superpotential terms. There is one term with $-1$ ($+1$) coefficient for each clock-wise (anti-clock-wise) closed triangle. The cubic terms in the superpotential are a remnant of the cubic $\mathcal{N}=4$ superpotential.
    \item $\mathcal{W}_{monopole}$ are the monopole terms in the superpotential generated by the Polyakov mechanism\cite{Polyakov_1977} due to the Higgsing of a $U(k)$ gauge symmetry to $U(1)^k$.
    For each vertical arrow, there is a monopole superpotential with 
    GNO flux $+1$ and $-1$ under the nodes connected by the arrow, from top to bottom.
\end{itemize}

\subsection{Checks of the Duality}

As a preliminary check of the proposed duality, we count the rank of the global symmetries of the planar theory. All $U(1)$ symmetries rotating the chiral fields are either broken by $\mathcal{W}_{planar}$ or removed by gauge transformations. The monopole superpotential breaks the topological $U(1)$ symmetries of the black nodes in each column to a diagonal combination. We thus have a single $U(1)$ topological symmetry for each column and an extra one associated with the node indicated in red, resulting in a $U(1)^{F}$ global symmetry.
The duality predicts that the UV global symmetry enhances in the IR to $U(F)$, which is manifest in the UV in the electric \SQCD. This is inherited from the topological symmetry enhancement in the $\mathcal{N}=4$ quiver of Figure \ref{fig:N4} %\eqref{quiv: 3d_N=4_mirror} 
which is preserved by the real mass deformation we consider.

In addition to the already mentioned check of the matching of the $\mathbf{S}^3_b$ partition functions, we also matched the refined Superconformal Indices (SCIs) %\cite{Romelsberger:2005, Imamura_2011, Kapustin_2011}  
for 
$N<5$ and $F<11$ up to $\mathcal{O}(x^{12/5})$, with trial R-charge of the baryons set to $R=\frac{3}{5}$. 
In particular, we can detect 
the character of the enhanced $U(F)$ current 
at order $x^2$ in the SCI of the planar Abelian mirror.

\subsection{Structure of the Moduli Space}

We observe that the chiral ring generators in the electric $SU(N)$ with $F$ fundamentals theory are the $\binom{F}{N}$ baryons $B^{j_1,\dots,j_N}=%\epsilon_{i_1\ldots i_{F-N}j_1\ldots j_{N}}
\epsilon^{a_1\ldots a_N}Q_{a_1}^{j_1}\ldots Q_{a_N}^{j_N}$\footnote{For $N=2$, $F=4$, $k=0$ there is also a single gauge invariant monopole $\mathfrak{M}$ on the electric side. This is mapped to a monopole in the mirror $ \hat{\mathfrak{M}}^{\left(\:\hspace{-2pt}
        \resizebox{20pt}{!}{%
        \begin{tabular}{ccc} 2&2&1\\1&0&
        \end{tabular}}\right)
}$ as well. This exception can be easily understood considering that  $SU(2)_0$ with $[4,0]$ flavors is the same as $SU(2)_0$ with $[2,2]$ flavors or $USp(2)_0$ with $4$ fundamentals which clearly has
a chiral ring monopole and the associated Coulomb branch.
For $N\geq3$, $F=2N$, $k=0$ from the SCI we detect the presence of a chiral dressed monopole. However, the monopole is nilpotent and there is no Coulomb branch \cite{Aharony_2015a}.
%, and, hence, are not part of the duality map.
}. The full set of BPS baryons can be encoded in the following Hilbert Series \cite{Feng_2006,Benvenuti_2006, Gray_2008}:%\SR{Add noppy and hanany}
\begin{equation}\label{eq:BPS_HS}
\begin{aligned}
    \mathcal{HS}_{baryons} &= \sum_{k=0}^{\infty} [0^{N-1},k,0^{F-N-1}]_{SU(F)}t^{k}\\ &\overset{(u.r.)}{\longrightarrow} \; \frac{\sum_{k=0}^Ac_{N,F}^{(k)} t^k}{(1-t)^d},
    \end{aligned}
\end{equation}
where $[\lambda_1, \ldots, \lambda_{F-1}]$ is the Dynkin label of the $SU(F)$ representation. We have turned off the fugacities and resummed the unrefined Hilbert Series in the second line of \eqref{eq:BPS_HS} (indicated by $u.r.$), wherein $A=F(N-1)-N^2+1$, $c_{N,F}^{(k)}=c_{N,F}^{(A-k)}>0$, $c_{N,F}^{(k)}=c_{F-N,F}^{(k)}$ and $d=FN-N^2+1$. 
We deduce that the branch of the moduli space generated by the baryons is a $d$-dimensional complex cone.

In the planar theory, the chiral ring is generated by monopoles, as we will see explicitly in an example below.
The proposed duality predicts that they satisfy quantum relations compatible with the Hilbert Series \eqref{eq:BPS_HS}.

\subsection{An Example: $SU(2)_{\frac{1}{2}}$ with 5 fundamentals}
Let us consider the example of $SU(2)_{\frac{1}{2}}$ \SQCD\, with 5 fundamental flavors:

\begin{equation}\label{su2w5}
\includegraphics[]{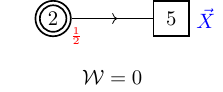}
\end{equation}
where $X_i$, $i=1,\dots,5$ are the real masses for the Cartan of the flavor $U(5)_{\vec{X}}$ symmetry and the chiral fields are assigned a trial R-charge $\frac{1}{2}$.
Our proposal for the mirror theory is given below:

\begin{equation}\label{generalmirrorsu2}
\includegraphics[]{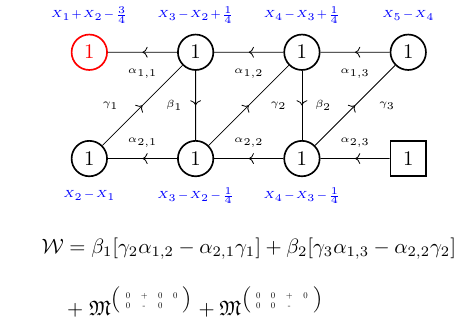}
\end{equation}
with the same convention for the CS and mixed CS levels as in Figure \ref{fig:N2}. 
The fields $\beta_i$ have trial R-charge $1$ and the fields $\alpha_{i,j},\gamma_k$ have trial R-charge $\frac{1}{2}$.
The labels in blue are the FI parameters for the corresponding node, namely the real masses for the topological symmetries. 
These are written in terms of the $X_i$, reflecting the embedding in the enhanced symmetry:
\begin{equation}
U(1)^5_{top}\times U(1)_R \to U(5)_{\vec{X}}\times U(1)_R 
\end{equation}
where $U(1)^5_{top}$ is the subgroup of the seven topological symmetries unbroken by the monopole superpotential.
Notice that the linear monopole superpotential induces a mixing between the topological and the trial R-symmetry, encoded in the constant terms in the FI terms \eqref{generalmirrorsu2}.

The chiral ring of the electric theory is generated by the $\binom{5}{2}=10$ baryons $B$. 
The mapping to the operators on the mirror side is:
\begin{equation}
    \begin{split}
        B \; \leftrightarrow & \Bigg \{  \mathfrak{M}^{\left(\:\hspace{-2pt}
        \resizebox{30pt}{!}{%
        \begin{tabular}{cccc} -&0&0&0\\0&0&0&
        \end{tabular}
        }\hspace{-2pt}\right)},\mathfrak{M}^{\left(\:\hspace{-2pt}
        \resizebox{30pt}{!}{%
        \begin{tabular}{cccc} -&-&0&0\\0&0&0&
        \end{tabular}
        }\hspace{-2pt}\right)},\mathfrak{M}^{\left(\:\hspace{-2pt}
        \resizebox{30pt}{!}{%
        \begin{tabular}{cccc} -&-&-&0\\0&0&0&
        \end{tabular}
        }\hspace{-2pt}\right)}, \\
        & \mathfrak{M}^{\left(\:\hspace{-2pt}
        \resizebox{30pt}{!}{%
        \begin{tabular}{cccc} -&-&-&-\\0&0&0&
        \end{tabular}
        }\hspace{-2pt}\right)},\mathfrak{M}^{\left(\:\hspace{-2pt}
        \resizebox{30pt}{!}{%
        \begin{tabular}{cccc} -&-&0&0\\-&0&0&
        \end{tabular}
        }\hspace{-2pt}\right)},\mathfrak{M}^{\left(\:\hspace{-2pt}
        \resizebox{30pt}{!}{%
        \begin{tabular}{cccc} -&-&-&0\\-&0&0&
        \end{tabular}
        }\hspace{-2pt}\right)}, \\
        & \mathfrak{M}^{\left(\:\hspace{-2pt}
        \resizebox{30pt}{!}{%
        \begin{tabular}{cccc} -&-&-&-\\-&0&0&
        \end{tabular}
        }\hspace{-2pt}\right)}, 
        \mathfrak{M}^{\left(\:\hspace{-2pt}
        \resizebox{30pt}{!}{%
        \begin{tabular}{cccc} -&-&-&0\\-&-&0&
        \end{tabular}
        }\hspace{-2pt}\right)}, 
        \mathfrak{M}^{\left(\:\hspace{-2pt}
        \resizebox{30pt}{!}{%
        \begin{tabular}{cccc} -&-&-&-\\-&-&0&
        \end{tabular}
        }\hspace{-2pt}\right)}, \\
        &
           \mathfrak{M}^{\left(\:\hspace{-2pt}
        \resizebox{30pt}{!}{%
        \begin{tabular}{ccccc} -&-&-&-\\-&-&-&
        \end{tabular}
        }\hspace{-2pt}\right)}
        \Bigg \}
    \end{split}
\end{equation}
which can be verified by computing the charges of monopole operators. 
The charges of a monopole with GNO fluxes $m_i$, $i=1,\dots,n_g$ under the $n_g$ Abelian gauge symmetries, can be compactly encoded in the following polynomial of the fugacities \cite{Aharony_1997,Benini:2011cma}:
\begin{equation}	\label{eq:monopole_charge}
\begin{split}
\mathcal{Q}&(\vec{m}) = -\frac{1}{2} \Bigg(
		\sum_{b_{ij}} |m_i - m_j| \left((R[b_{ij}]-1) - u_i + u_j  \right)
		\\&\qquad+
		\sum_{f_{i},\tilde{f}_{i}} |m_i| \left( (R[\tilde{f}_{i}, f_i]-1) \pm u_i  \right)
	\Bigg)
\\&      \qquad +\sum_{i=1}^{n_g} \lambda_i m_i
  -\frac{1}{2} \sum_{i\leq j} k_{ij}(m_i u_j + m_j u_i),
	%-\sum_{i=1}^{n_g} k_i m_i u_i
	%+\sum_{i=1}^{n_g} \lambda_i m_i
	%-\frac{1}{2}\sum_{i,j} k_{ij} (m_i u_j + m_j u_i)
\end{split}
\end{equation}

where the first two lines are contributions from fermionic zero modes in the bifundamentals $b_{ij}$ and (anti)fundamentals ($\tilde{f_i}$) $f_i$ and the other terms are due to CS levels $k_i$, mixed CS levels $k_{ij}$ and FI terms $\lambda_i$.
In our conventions, the CS and FI terms are given by (the SUSY completion of):
\begin{equation}
    -\sum_{l,m}i \frac{k_{lm}}{4 \pi } \int A_l \wedge dA_m +\sum_l  \lambda_l \int D_l,
\end{equation}
where $A_j$  is the gauge field of the $j-$th node, and with $k_{jj}=k_j$ being the CS level for the $j-$th node, while the FI terms $\lambda_i$ come from mixed CS interactions between the gauge group $U(1)_i$ and the associated topological symmetry $U(1)_{T_i}$ and $D_i$ is the auxiliary field in the vector superfield containing the gauge field $A_i$. 

In Equation \eqref{eq:monopole_charge}, the charge under a gauge or flavor symmetry is encoded in the coefficient in front of the corresponding parameter, $u_i$ are fugacities for the gauge symmetries and the constant is the trial R-charge.
One can check that the charge of the monopoles in the superpotential and the chiral ring map do not depend on the $u_i$ and they are, therefore, gauge invariant. In particular:

\begin{equation}
\mathcal{Q} \left[
    \mathfrak{M}^{\left(\:\hspace{-2pt}
        \resizebox{30pt}{!}{%
        \begin{tabular}{cccc} 0&+&0&0\\0&-&0&
        \end{tabular}}\right)
}\right]
=
\mathcal{Q} \left[
    \mathfrak{M}^{\left(\:\hspace{-2pt}
        \resizebox{30pt}{!}{%
        \begin{tabular}{cccc} 0&0&+&0\\0&0&-&
        \end{tabular}}\right)
}\right] =2
\end{equation}
consistent with the monopole superpotential and:
\begin{equation}
\begin{array}{l}
\mathcal{Q} \left[
    \mathfrak{M}^{\left(\:\hspace{-2pt}
        \resizebox{30pt}{!}{%
        \begin{tabular}{cccc} -&0&0&0\\0&0&0&
        \end{tabular}}\right)
}\right]
= \mathcal{Q} \left[ B^{1,2} \right]
= 1 - X_1 - X_2
\\
\vdots
\\
\mathcal{Q} \left[
    \mathfrak{M}^{\left(\:\hspace{-2pt}
        \resizebox{30pt}{!}{%
        \begin{tabular}{cccc} -&-&-&-\\-&-&-&
        \end{tabular}}\right)
}\right]
= \mathcal{Q} \left[   B^{4,5} \right]
= 1 - X_4 - X_5
\end{array}
\end{equation}
compatible with the chiral ring map. 

We also report the superconformal index of both theories:
\begin{equation}\label{sciexpansion}
\mathcal{I}=1+\mathbf{10}fx^{3/5}+\mathbf{50}f^2 x^{6/5}+\mathbf{175}f^3 x^{9/5}-(\mathbf{24}+1)x^2+\mathcal{O}(x^{12/5})
\end{equation}
where bold numbers denote $SU(5)$ representations and $f$ is the fugacity for the baryonic $U(1)\subset U(5)$. 
Notice that to perform the SCI expansion we shifted the trial R-charge of the fundamentals in the \SQCD\, so that baryons have R-charge $\frac{3}{5}$, which is closer to the superconformal value that can be computed via F-extremization \cite{Jafferis_2012} ($R_{sc}=0.67778\dots$). The shift can be performed by changing the mixing coefficient of the $U(1)$ baryonic symmetry. In the mirror theory in \eqref{generalmirrorsu2}, this corresponds to shifting only the FI terms of the red gauge node by a constant number.
In the index in \eqref{sciexpansion}, we observe terms of the form $[0,k,0,0]_{SU(5)}f^k x^{k R }$, corresponding to the baryons on the electric side and to monopoles on the mirror side, and the negative term at order $x^2$ corresponding to the $U(5)$ conserved currents.

%%%%%%%%%%%%%%%%%%%%%%%%%%%%%%%%%%
\section{More chiral $\leftrightarrow$ planar Duals}
Our analysis can be extended to obtain similar Abelian planar duals for unitary \SQCD\; with both fundamental and anti-fundamental matter and more general CS levels. As will be discussed in forthcoming papers \cite{Benvenuti_2024, Benvenuti_2024a}, it is also possible to 
define an algorithmic procedure in the spirit of \cite{rethinking,sl2z,Benvenuti2_2024}, which streamlines the study of unitary linear and circular chiral quiver gauge theories. 

This section discusses an example of an \SQCD\, with a more general CS level obtained via further real mass deformations and a linear quiver gauge theory.

Abelian mirror duals of $3d$ $\mathcal{N}=2$ non-Abelian quivers appeared in \cite{Benvenuti:2017lle, Benvenuti:2017kud, Benvenuti:2017bpg}; a key difference in \cite{Benvenuti:2017lle, Benvenuti:2017kud, Benvenuti:2017bpg} from our proposal is that the non-Abelian side is non-chiral and the Abelian side is linear instead of planar.

\subsection{Real Mass Deformations}\label{subsec:mass}

The procedure outlined in the previous section allows for the construction of Abelian planar duals for $SU(N)$ SQCD with CS level $k=\frac{F}{2}-N\geq0$. We can generalize to $k>\frac{F}{2}-N$ by turning on further positive real mass deformations for fundamental fields. As an example, we consider the duality for $SU(2)$ with $5$ fundamentals \eqref{su2w5} studied in the previous section, and turn on a real positive mass associated to $X_5$, making one fundamental massive\footnote{Generically, real masses for the other fundamental fields correspond to different mirror dual frames for \eqref{eq:SU2_4fund_massdef}.} and flowing to:
\begin{equation}    \label{eq:SU2_4fund_massdef}
\includegraphics[]{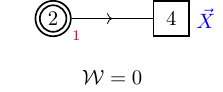}
\end{equation}

We claim that the corresponding vacua on the mirror side \eqref{generalmirrorsu2}
is such that the fields $\alpha_{1,3},\gamma_3,\alpha_{2,3}$
are massive. 
This can be followed on the $\mathbf{S}_b^3$ partition function, where the matching of asymptotics in the large $X_5$ limit provides a non-trivial check of the flow.
%the real mass flow renders some of the fields massive and we integrate them out. 
Notice that all the fields charged under the rightmost gauge node are massive, leaving a $U(1)_{-1}$ CS theory coupled to the rest of the quiver by BF terms. 
Furthermore, by a gauge field redefinition, we can render all chiral fields neutral under the gauge symmetry of the bottom right node. Then, the corresponding gauge field describes a decoupled $U(1)_{-1}$ sector, and the bottom right node effectively becomes a flavor node.

The path integral over the two $U(1)_{-1}$ gauge fields can be performed exactly\cite{Kapustin:1999ha, Witten:2003ya}, resulting in:
%\begin{equation}\label{generalmirrorsu2_massdef}
\begin{eqnarray}\label{generalmirrorsu2_massdef}
\makebox[0.8\linewidth]{
\includegraphics[]{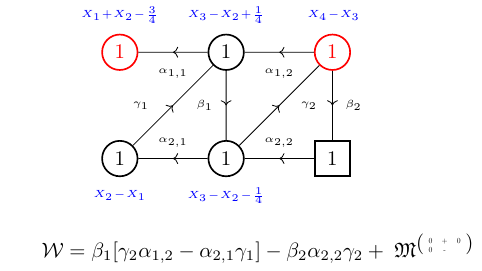}
} %end of makebox
\end{eqnarray}
%\end{equation}
Notice that there are two gauge nodes at CS level $-\frac{1}{2}$ (indicated in red).
The duality map between the baryons on the electric side and the %strings of
monopoles on the mirror side is now given by:
\begin{equation}
    \begin{split}
        B \; \leftrightarrow & \Bigg \{  \mathfrak{M}^{\left(\:\hspace{-2pt}
        \resizebox{20pt}{!}{%
        \begin{tabular}{ccc} -&0&0\\0&0&
        \end{tabular}
        }\hspace{-2pt}\right)},
        \mathfrak{M}^{\left(\:\hspace{-2pt}
        \resizebox{20pt}{!}{%
        \begin{tabular}{ccc} -&-&0\\0&0&
        \end{tabular}
        }\hspace{-2pt}\right)},
        \mathfrak{M}^{\left(\:\hspace{-2pt}
        \resizebox{20pt}{!}{%
        \begin{tabular}{ccc} -&-&-\\0&0&
        \end{tabular}
        }\hspace{-2pt}\right)},
        \\&
        \mathfrak{M}^{\left(\:\hspace{-2pt}
        \resizebox{20pt}{!}{%
        \begin{tabular}{ccc} -&-&0\\-&0&
        \end{tabular}
        }\hspace{-2pt}\right)},
        \mathfrak{M}^{\left(\:\hspace{-2pt}
        \resizebox{20pt}{!}{%
        \begin{tabular}{ccc} -&-&-\\-&0
        \end{tabular}
        }\hspace{-2pt}\right)},
        \mathfrak{M}^{\left(\:\hspace{-2pt}
        \resizebox{20pt}{!}{%
        \begin{tabular}{ccc} -&-&-\\-&-&
        \end{tabular}
        }\hspace{-2pt}\right)}
        \Bigg \}
    \end{split}
\end{equation}
which can be checked by computing the charges of monopole operators as described in the previous section.

\subsection{A Linear Quiver with Chiral Matter and its Planar Abelian Dual}\label{sec:tsun}

We can perform similar real mass deformations in mirror pairs of unitary quiver $\mathcal{N}=4$ theories.
Here we consider the example of the $T[SU(N)]$ theory\cite{Gaiotto_2008}.
The global symmetry is $SU(N)_{\vec{X}} \times SU(N)_{\vec{Y}}$ 
and mirror symmetry is a self-duality that exchanges the two $SU(N)$ symmetries. 
We turn on a real mass deformation for the commutant of $U(1)_R$ and the $SU(N)_{\vec{Y}}$ symmetry that breaks SUSY to $\mathcal{N}=2$ and breaks $SU(N)_{\vec{Y}} \to U(1)_{Y_i}^{N-1}$. We choose a vacuum in which the chiral adjoint multiplets and half of the chiral bifundamental fields are integrated. The dual vacuum in the mirror side Abelianizes all the gauge groups. For $N=3$, we find the following duality: 

\begin{equation}    \label{eq:GUN_quivers}
\includegraphics[scale=0.9]{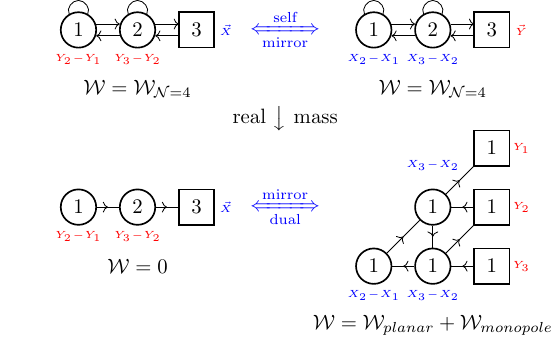}
\end{equation}

In the chiral linear quiver, 
there is a CS term at level 1 for the diagonal $U(1)\subset U(m)$ for each $U(m)$ gauge group and there is a mixed CS term at level $-1$ between adjacent nodes.
The CS and mixed CS couplings of the mirror planar Abelian quiver follow from the prescription outlined in Section \ref{sec:1ststep}.

The chiral rings of the two theories also match, with the $N-1$ dressed gauge invariant monopoles of the electric theory mapped to the $N-1$ mesonic operators constructed along the shortest path connecting two adjacent $U(1)$ flavor nodes.

%%%%%%%%%%%%%%%%%%%%%%%%%%%%%%%%%%%%%%%%%%%%%%%
\acknowledgments
We are grateful to Amihay Hanany for useful conversations. SB and SP are partially supported by the MUR-PRIN grant No. 2022NY2MXY. SR is supported by the MUR-PRIN grant No. 2022NY2MXY.

\end{document}